\documentclass[twocolumn,english,aps,prb,twocolum,superscriptaddress,bibnotes,amsmath,amssymb,floatfix]{revtex4-1}
\usepackage[colorlinks=true,citecolor=blue,linkcolor=magenta]{hyperref}

\usepackage[markup=nocolor, authormarkupposition=left]{changes} 
\usepackage{soul}
\usepackage[utf8]{inputenc}
\usepackage[english]{babel}
\usepackage{amsmath,amsfonts,amssymb}
\usepackage[T1]{fontenc}
\usepackage{url}

\usepackage{amsmath}
\usepackage{amsfonts}
\usepackage{amssymb}

\usepackage{epstopdf}
\usepackage{graphicx}
\graphicspath{{./Figures/}}

\usepackage{changes}

\begin{document}

\title{Programmable access to microresonator solitons with modulational sideband heating}

\author{Huamin Zheng}
\affiliation{College of Electronics and Information Engineering, Shenzhen University, Shenzhen 518000, China}
\affiliation{International Quantum Academy, Shenzhen 518048, China}

\author{Wei Sun}
\email[]{sunwei@iqasz.cn}
\affiliation{International Quantum Academy, Shenzhen 518048, China}

\author{Xingxing Ding}
\affiliation{Key Laboratory of Specialty Fiber Optics and Optical Access Networks, Shanghai University, Shanghai 200444, China}

\author{Haoran Wen}
\affiliation{International Quantum Academy, Shenzhen 518048, China}
\affiliation{School of Science and Engineering, CUHK(SZ), Shenzhen 518100, China}

\author{Ruiyang Chen}
\affiliation{International Quantum Academy, Shenzhen 518048, China}
\affiliation{Shenzhen Institute for Quantum Science and Engineering, Southern University of Science and Technology, Shenzhen 518055, China}

\author{Baoqi Shi}

\affiliation{International Quantum Academy, Shenzhen 518048, China}
\affiliation{Department of Optics and Optical Engineering, University of Science and Technology of China, Hefei 230026, China}

\author{Yi-Han Luo}
\affiliation{International Quantum Academy, Shenzhen 518048, China}
\affiliation{Shenzhen Institute for Quantum Science and Engineering, Southern University of Science and Technology, Shenzhen 518055, China}

\author{Jinbao Long}
\affiliation{International Quantum Academy, Shenzhen 518048, China}

\author{Chen Shen}
\affiliation{International Quantum Academy, Shenzhen 518048, China}

\author{Shan Meng}
\affiliation{College of Electronics and Information Engineering, Shenzhen University, Shenzhen 518000, China}

\author{Hairun Guo}
\affiliation{Key Laboratory of Specialty Fiber Optics and Optical Access Networks, Shanghai University, Shanghai 200444, China}

\author{Junqiu Liu}
\email[]{liujq@iqasz.cn}
\affiliation{International Quantum Academy, Shenzhen 518048, China}
\affiliation{Hefei National Laboratory, University of Science and Technology of China, Hefei 230088, China}

\maketitle

\noindent\textbf{Dissipative Kerr solitons formed in high-$Q$ optical microresonators provide a route to miniaturized optical frequency combs that can revolutionize precision measurements, spectroscopy, sensing, and communication.  
In the last decade, a myriad of integrated material platforms have been extensively studied and developed to create photonic-chip-based soliton combs. 
However, the photo-thermal effect in integrated optical microresonators has been a major issue preventing simple and reliable soliton generation. 
Several sophisticated techniques to circumvent the photo-thermal effect have been developed. 
In addition, instead of the single-soliton state, emerging applications in microwave photonics and frequency metrology prefer multi-soliton states. 
Here we demonstrate an approach to manage the photo-thermal effect and facilitate soliton generation. 
The approach is based on a single phase-modulated pump, where the generated blue-detuned sideband synergizes with the carrier and thermally stabilizes the microresonator. 
We apply this technique and demonstrate deterministic soliton generation of 19.97 GHz repetition rate in an integrated silicon nitride microresonator. 
Furthermore, we develop a program to automatically address to target $N-$soliton state, in addition to the single-soliton state, with near 100\% success rate and as short as 10 s time consumption.  
Our method is valuable for soliton generation in essentially any platforms even with strong photo-thermal effect, and can promote wider applications of soliton frequency comb systems for microwave photonics, telecommunication and frequency metrology. 
}

Dissipative Kerr solitons formed in high-$Q$ optical microresonators \cite{Herr:14, Kippenberg:18} constitute miniaturized optical frequency combs with broad bandwidths and repetition rates in the microwave to millimeter-wave domain. 
Commonly referred to as ``soliton microcombs'', they have been already used in many system-level information and metrology applications, such as coherent telecommunication \cite{Marin-Palomo:17, Corcoran:20, Mazur:21}, ultrafast ranging \cite{Trocha:18, Suh:18}, astronomical spectrometer calibration \cite{Obrzud:19, Suh:19}, dual-comb spectroscopy \cite{Dutt:18, Yang:19}, low-noise microwave generation \cite{Liang:15, Liu:20}, photonic neural networks \cite{Feldmann:21, Xu:21}, datacenter circuit switch \cite{Raja:21}, microwave photonics \cite{Wu:18, Hu:20}, frequency synthesizers \cite{Spencer:18}, and optical atomic clocks \cite{Newman:19}. 
Critical to the rapid progress of soliton microcomb technology is the development and continuous optimization of various photonic integrated platforms including silica \cite{Yi:15, Yang:18}, silicon nitride (Si$_3$N$_4$) \cite{Brasch:15, Joshi:16}, high-index doped silica (Hydex) \cite{Bao:19, WangX:21}, aluminium nitride (AlN) \cite{LiuX:21, WengH:21}, lithium niobate (LiNbO$_3$) \cite{He:19, Gong:20}, tantala pentaoxide (Ta$_2$O$_5$) \cite{Jung:21}, silicon carbide (SiC) \cite{Guidry:21, Wang:22}, chalcogenide \cite{XiaD:22}, aluminium gallium arsenide (AlGaAs) \cite{Pu:16, Moille:20} and gallium phosphide (GaP) \cite{Nardi:23}. 
Besides, the introduction and successful implementation of hybrid and heterogeneous integration \cite{Stern:18, Liu:20a, Voloshin:21, Xiang:21, Churaev:23, LiuY:22, OpdeBeeck:20} further enable complex control schemes, extra nonlinearity and efficient amplification for integrated soliton microcombs. 

Despite these advances, one issue that currently prevents a wider deployment of soliton microcombs is to deterministically access and control the soliton state. 
This issue is caused by the photo-thermal effect, arising from light absorption and thermal-optic effect in optical microresonators \cite{Li:17, Gao:22}, particularly for those built on integrated platforms. 
When the CW pump's frequency is scanned through a microresonator resonance, from the blue-detuned side to the red-detuned side, a self-organized pulse waveform (i.e. a soliton state), is formed \cite{Herr:14}. 
However, the photo-thermal effect leads to serious thermal instability that often annihilates the soliton state immediately.  
Therefore, sophisticated techniques to manage this effect have been developed, such as power kicking \cite{Brasch:16, Yi:16b}, single-sideband suppressed-carrier frequency shifters \cite{Stone:18}, dual-laser pump \cite{Zhou:19, ZhangS:19, Drake:20}, pump modulation \cite{Jang:15, Wildi:19, Nishimoto:22, Cole:18, Miao:22}, pulse pumping \cite{Obrzud:17}, or laser self-cooling \cite{Lei:22,Grassani:22}. 
Besides, cryogenic operation can be helpful \cite{Moille:20}. 
In addition, instead of the single-soliton state, efforts have been also made on employing multi-soliton states for reconfigurable photonic microwave filters \cite{Hu:20} and synthesis of terahertz frequency \cite{LuZ:21}. 
For frequency metrology applications, the local comb line power enhancement in multi-soliton state further facilitates the realization of self-referencing \cite{Brasch:17}.

\begin{figure*}[t!]
\centering
\includegraphics{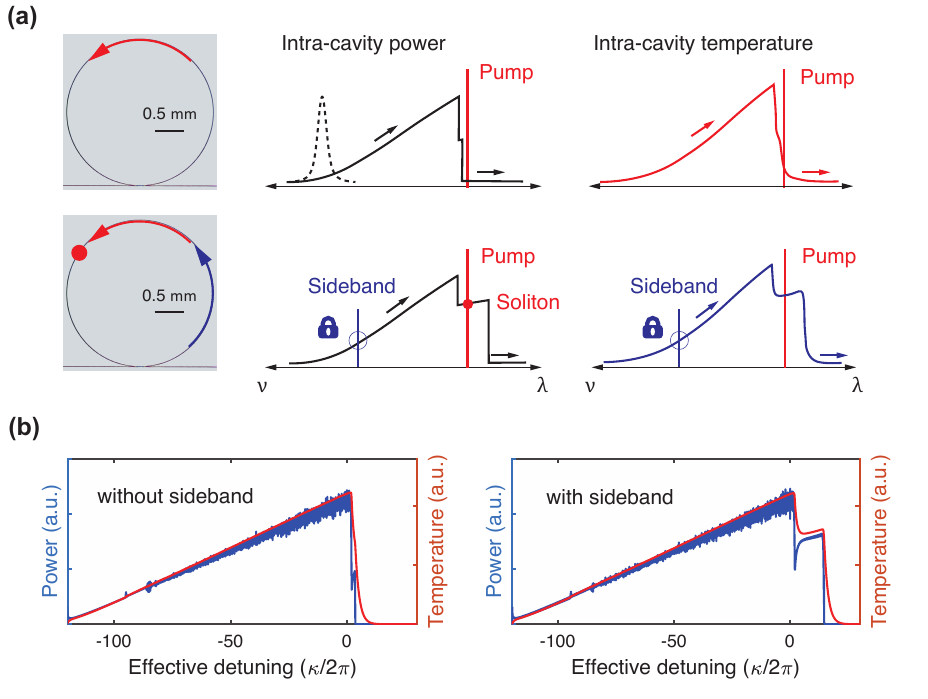}
\caption{
\textbf{Principle and numerical simulation of soliton generation in optical microresonators assisted with sideband heating.}
\textbf{(a)}. 
Comparison of soliton step length without (top row) and with (bottom row) sideband heating. 
Left column shows the photos of the same Si$_3$N$_4$ microresonator of 19.97 GHz FSR. 
Red and blue arc arrows represent the CW carrier pump and the blue-detuned sideband. 
Middle column show the intra-cavity power dynamics (solid black line) and the ``cold'' resonance (dashed black line). 
Right column shows the intra-cavity temperature dynamics. 
If the scanning pump (red line) stops on the soliton step, a soliton state is formed (red spot). 
Such a process can be facilitated by the blue-detuned sideband (blue line), which locks the ``hot'' resonance, avoid its shift, and extend the soliton step.
\textbf{(b)}. 
Numerical simulation of the intra-cavity power (blue) and temperature(red), without (left) and with (right) the blue-detuned sideband.
The simulation reveals soliton step extension due to the blue-detuned sideband. 
}
\label{Fig:1}
\end{figure*}

Here we demonstrate an approach to manage the photo-thermal effect, and to automatically and deterministically access to a soliton state.
This method, employing single-sideband heating, overcomes the photo-thermal effect in integrated microresonators. 
We apply this method to demonstrate that, $N-$soliton states of 19.97 GHz repetition rate can be deterministically generated for any assigned value of $N\leq5$. 

\noindent\textbf{Soliton step extension via sideband heating}.
The principle to excite a soliton state in a high-$Q$ optical microresonator with a single continuous-wave (CW) pump laser is shown in Fig.~\ref{Fig:1}(a) top.
As the pump laser scans through a resonance from its blue-detuned to red-detuned side, the resonance is distorted from a Lorentzian to a triangular profile \cite{Carmon:04}. 
This change is due to the increasing intra-cavity power, which, via the the photo-thermal effect, levitates the microresonator's temperature and extends its total optical length. 
Therefore, the ``hot'' resonance shifts following the direction of laser scanning. 
In this case, solitons can form at the red-detuned side of the effective resonance. 
A signature indicating soliton states is a series of step features (``soliton step'') in the intra-cavity power \cite{Guo:16} as shown in Fig.~\ref{Fig:1}(a). 
However, once a soliton state is reached, the intra-cavity power, as well as the microresonator temperature, drops suddenly. 
This leads to a sudden blue-shift of the resonance, which in return increases the effective soliton-pump detuning. 
Therefore, this transition often compromises the soliton stability, causes extremely short or even negligible soliton steps, and often annihilates solitons immediately. 

\begin{figure*}[t!]
\centering
\includegraphics{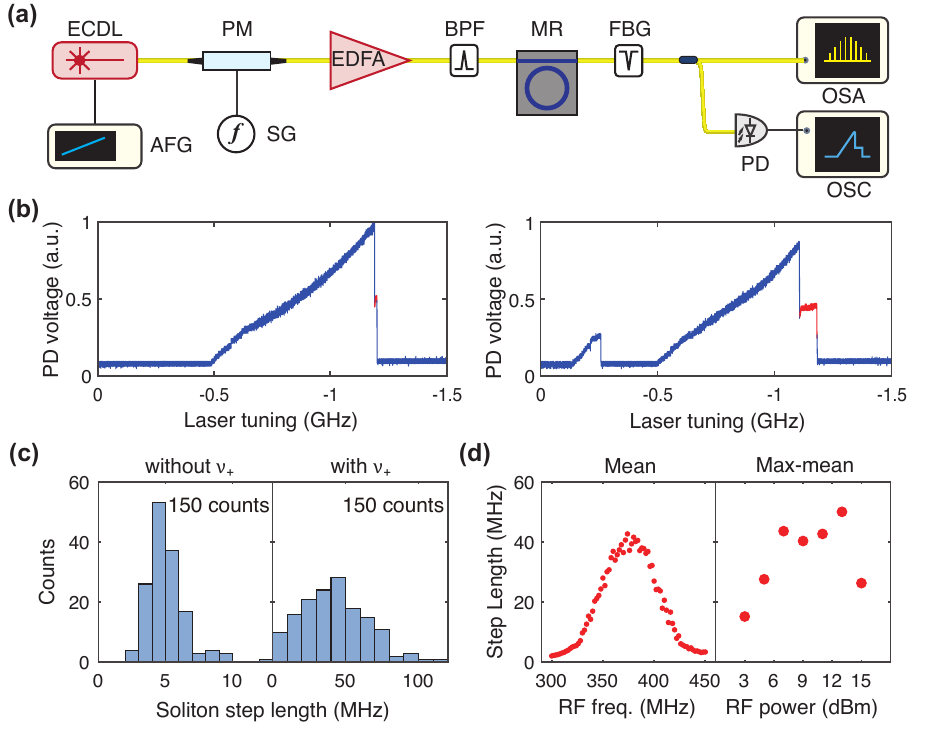}
\caption{
\textbf{Experimental demonstration of soliton step extension using sideband heating.}
\textbf{(a)}. 
Experimental setup. 
ECDL, external-cavity diode laser. 
AFG, arbitrary function generator. 
PM, phase modulator. 
SG, signal generator. 
EDFA, erbium-doped fiber amplifier. 
BPF, bandpass filter. 
MR, Si$_3$N$_4$ microresonator. 
FBG, fiber Bragg grating. 
PD, photodiode. 
OSA, optical spectrum analyzer. 
OSC, oscilloscope. 
\textbf{(b)}. 
Experimentally detected light generation from the microresonator, showing soliton step extension with the sideband. 
$\nu_0$, pump laser frequency. 
$\nu_+$, blue-detuend sideband frequency. 
$\nu_-$, red-detuend sideband frequency, which is idle in soliton generation. 
$\nu_\pm$ are generated by phase-modulating the pump laser with 396 MHz RF frequency and 11 dBm RF power on phase modulator. 
The arrows mark that the laser frequency is decreasing. 
The red-colored segments are soliton steps. 
In the right panel, with the sideband, the small triangular peak is due to $\nu_-$, the major triangular peak is due to $\nu_0$, and the soliton step is extended due to $\nu_+$. 
\textbf{(c)}. 
Statistical analysis of 150 measured soliton step length values without (left) and with (right) the sideband $\nu_+$. 
The mean soliton step length is increased from 5.2 MHz (left) to 42.7 MHz (right).
\textbf{(d)}. 
Left: With 11 dBm RF power, the dependence of soliton step length on RF frequency. 
Right: The maximum value of step length for a given RF power with varying RF frequency. 
The best combination found in our experiment is 352 MHz RF frequency and 13 dBm RF power. 
}
\label{Fig:2}
\end{figure*}

\begin{figure*}[t!]
\centering
\includegraphics{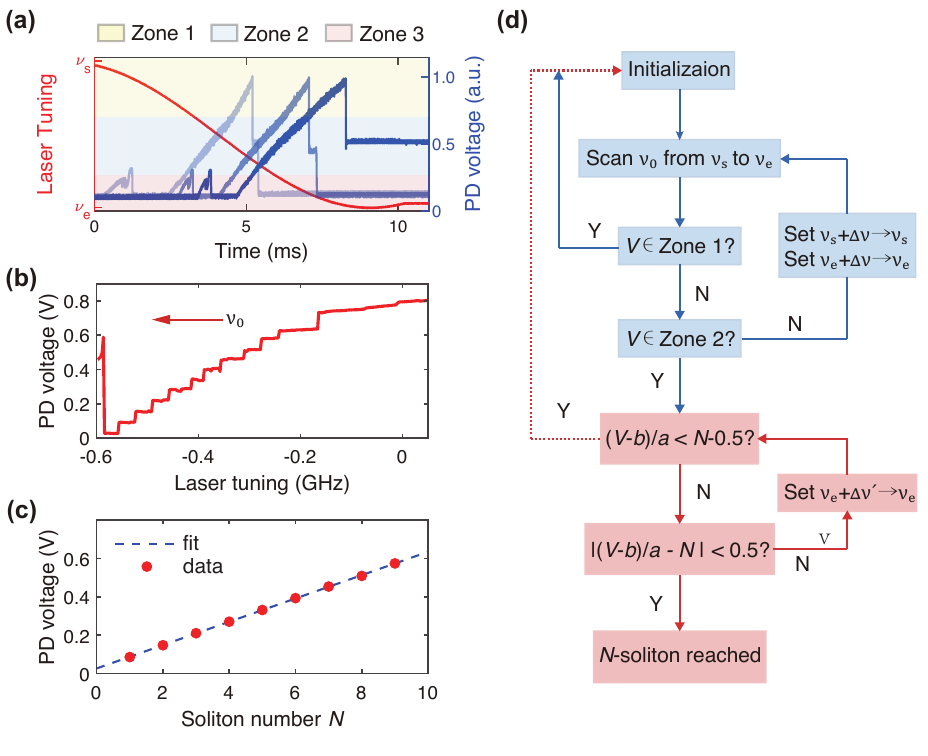}
\caption{
\textbf{Criteria and program for automatic $N-$soliton state addressing.}
\textbf{(a)}. 
Generated light power dynamics labeled with different zones. 
The red curve shows that the laser frequency $\nu_{0}$ is decreasing from $\nu_{s}$ to $\nu_{e}$.
The blue curves with different tones show the normalized power traces detected by the PD and recorded by the OSC. 
The traces together show the process to access a soliton state by translating $(\nu_{s}, \nu_{e})$ with step $\Delta \nu$ as shown in Panel (c).
Zone 1/2/3 are marked, and only Zone 2 supports soliton formation. 
\textbf{(b)}. 
Using the backward-tuning process, the power trace measured by the PD shows a stair feature, which contains many steps corresponding to soliton states with different soliton number $N$. 
The arrow marks that the laser frequency is increasing.
\textbf{(c)}.
The backward-tuning soliton generation process is repeated 20 times, and the average PD voltage $\overline{V}$ for each step are fitted with the soliton number $N$, resulting in a linear relation of $\overline{V}=a\times N +b$ ($a=0.0609, b=0.0265$).
\textbf{(d)}. 
The logic codes and algorithm flowchart for computer-controlled automatic $N-$soliton state generation. 
The blue-colored block is for soliton generation, and the red-colored block is for $N-$soliton state addressing. 
The solid lines with arrows represents successful addressing and the dashed line with arrows represents failed addressing.
}
\label{Fig:3}
\end{figure*}

To circumvent such thermally induced instability, here we develop a blue-detuned sideband heating scheme to balance the sudden cooling of the microresonator during soliton formation. 
The principle is depicted in Fig.~\ref{Fig:1}(a) bottom. 
With proper frequency and power differences between the pump laser and its sideband, wherever the resonance moves, the sideband will drag the resonance back. 
For example, when the resonance experiences a sudden blue-shift during the transition from modulation instability to a soliton state, more optical power of the sideband enters into the resonance, which levitates the microresonator's temperature to counteract the cooling. 
In such a way, the microresonator's intra-cavity power and temperature are locked, and the soliton step is extensively extended, facilitating soliton formation.  

A modified Lugiato-Lefever Equation (LLE) model to simulate soliton formation is considered as below,
\begin{equation}
    \begin{aligned}
        \frac\partial{\partial t}A(t,\varphi)&=-\frac \kappa2A(t,\varphi)-i[\delta_\omega+g_\omega|A|^2+\mathcal{D}_\omega+T]A(t,\varphi)\\
        &+\sqrt{\kappa_\text{ex}}A_\text{in},
    \end{aligned}
\end{equation}
where $A(t,\phi)$ is the intra-cavity field amplitude, 
$\kappa = \kappa_0 + \kappa_\text{ex}$ is the total optical loss rate (with $\kappa_0$ being the intrinsic loss rate and $\kappa_\text{ex}$ being the external coupling rate with the input pump laser), 
$\delta_\omega$ is the soliton detuning, 
$g_\omega$ is the nonlinear coefficient, 
$\mathcal{D}_\omega$ is the microresonator dispersion, 
$T$ is the microresonator temperature, 
$A_\text{in}=\sqrt{P_\text{in}}(1+\delta_\text{m}\text{exp}(+i2\pi f_\text{m}t))$ is the input laser amplitude,
$P_\text{in}$ is the input pump laser power, 
$P_\text{in}\delta_\text{m}^2$ is the input sideband power, 
$f_\text{m}$ is the frequency difference between the pump laser and the sideband.
The microresonator temperature $T$ evolves following the equation
\begin{equation}
    \frac{\partial T}{\partial t} =\frac{KT}{\tau_\text{r}}|A(t,\varphi)|^2-\frac {T}{\tau_\text{r}},
\end{equation}
where $KT$ is the thermal induced resonance shift and $\tau_\text{r}$ is the thermal relaxation time. 
With a proper set of parameters (see Supplementary Information), the simulation shows prominent efficacy for thermal stabilization and soliton step extension, as shown in Fig.~\ref{Fig:1}(b). 

Experimentally, we create the sideband by phase-modulation of the carrier pump laser. 
The experimental setup is shown in Fig.~\ref{Fig:2}(a).
The pump laser is an external-cavity diode laser (ECDL, Toptica CTL 1550) with its frequency $\nu_{0}$ tuned by a piezo which is actuated by an arbitrary function generator (AFG, Rigol DG4062). 
A phase modulator (PM, iXblue LPZ-LN-10) creates two sidebands $\nu_{-}$ and $\nu_{+}$ from the carrier pump laser. 
The modulated laser is then power-amplified by an erbium-doped fiber amplifier (EDFA, KEOPSYS CEFA-C-PB-HP). 
A band-pass filter is applied to reject amplifier spontaneous emission (ASE) noise from the EDFA. 
After that, the laser power is coupled into an integrated silicon nitride (Si$_3$N$_4$) optical microresonator fabricated using a DUV subtractive process \cite{Ye:23}. 
The Si$_3$N$_4$ microresonator features 19.97 GHz free spectral range (FSR) and an intrinsic quality factor of $Q_0\approx 1 \times 10^7$, characterized using a vector spectrum analyzer \cite{Luo:23} (see Supplementary Information). 
After the Si$_3$N$_4$ microresonator, the pump laser is blocked by two fiber Bragg gratings (FBG) providing 30 dB total attenuation. 
Finally, the output optical spectrum is monitored by an optical spectrum analyzer (OSA, Yokogava AQ6370D). 
The optical power is detected by a photodiode (PD, Thorlabs PDA10CS2) and monitored by an oscilloscope (OSC, Rigol DHO4404).

\begin{figure*}[t!]
\centering
\includegraphics{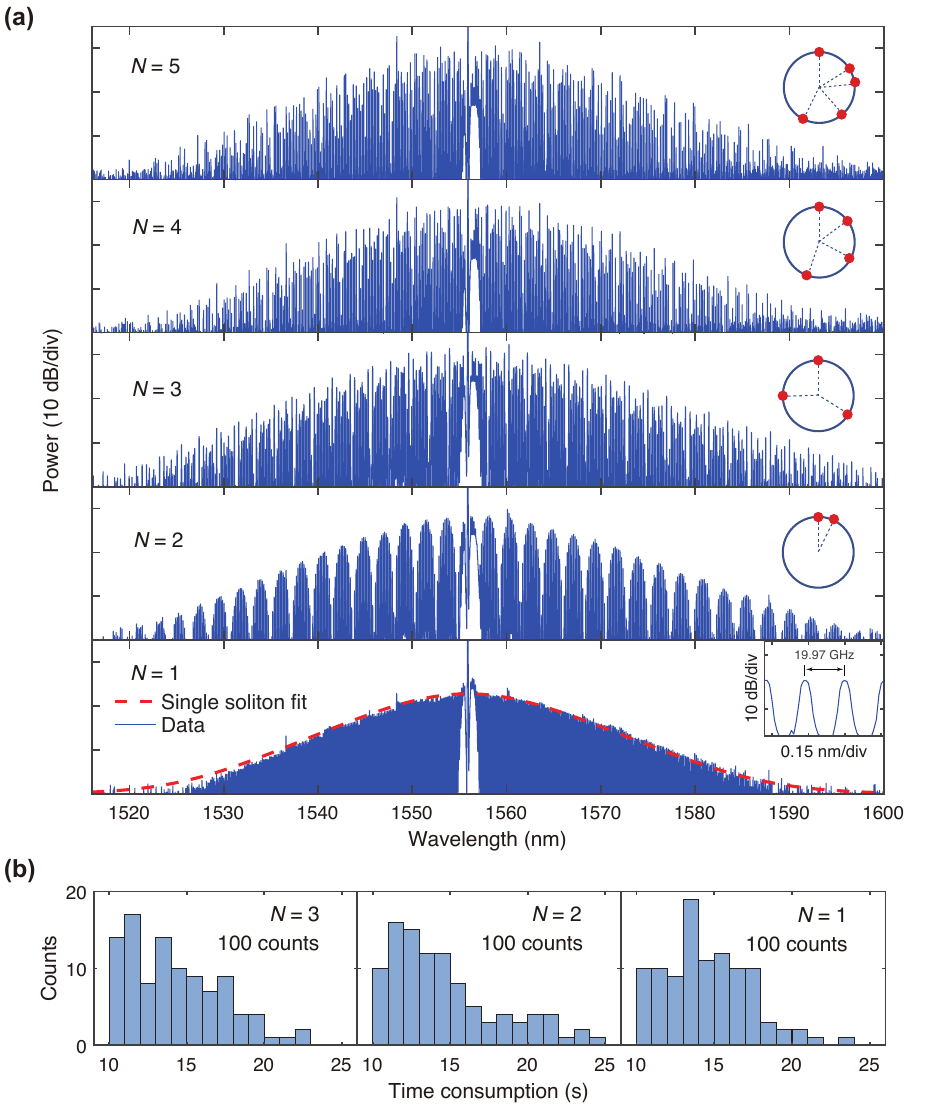}
\caption{
\textbf{single- and multi-soliton spectra, and the time consumption to reach these states.}
\textbf{(a)}. 
Optical spectra for $N=5,4,3,2,1-$soliton states of 19.97 GHz mode spacing. 
Insets show the relative soliton positions of the $N-$soliton state in the microresonator. 
The soliton numbers $N$ are retrieved by inverse Fourier transform on the spectrum. 
The single-soliton spectrum is fitted with a sech$^2$ function (dashed red line).
\textbf{(b)}. 
Statistics of time consumption for $N=3,2,1-$soliton state addressing with 100 repeated processes among which 93/99/100 processes come into successful state addressing.
}
\label{Fig:4}
\end{figure*}

The wavelength and power of the pump laser without phase modulation are $1555.7$ nm and 730 mW, respectively. 
The pump laser's frequency scans through a resonance from the effectively blue-detuned to red-detuned side (``forward tunining'' \cite{Herr:14}), with the direction marked by arrows in Fig.~\ref{Fig:2}(b). 
The light detected by the PD and recorded by the OSC corresponds to the generated light from Kerr parametric oscillation in the microresonator.
The soliton step is marked with red color within the total 1.5 GHz frequency range. 
Due to the photo-thermal effect in our Si$_3$N$_4$, the soliton step is so short that is barely seen, impossible to access a soliton state. 
However, when phase-modulating the pump laser with modulation frequency of $f_\text{m}=374$ MHz and RF power applied on the PM of 11 dBm, a pair of sidebands $\nu_{\pm}=\nu_{0} \pm f_\text{m}$ are created. 
The carrier pump laser's power is $8.3$ times higher than that of one sideband. 
Only the blue sideband $\nu_{+}$ is used to counteract the thermal effect \cite{Wildi:19, Nishimoto:22}, while the red sideband $\nu_{-}$ is idle. 
Prominent soliton step extension is observed with the phase modulation, as shown in Fig.~\ref{Fig:2}(b).
We repeatedly scan the laser through the resonance for 150 times, and statistically analyze the soliton step length as shown in Fig.~\ref{Fig:2}(c). 
The statistical mean values are 5.2 and 42.7 MHz, respectively, corresponding to 8-times extension due to sideband heating. 

We further investigate the optimal value of $f_\text{m}$ for maximum soliton step extension, as shown in Fig.~\ref{Fig:2}(d) left. 
The optimal value under 11 dBm RF power is $f_\text{m}=374$ MHz. 
The effect of RF power on soliton step length is also investigated as shown in Fig.~\ref{Fig:2}(d) right.

\noindent\textbf{Soliton number-state addressing}.
Next, we realize deterministic generation of $N-$soliton state using the sideband heating method. 
Figure~\ref{Fig:3}(a) shows the dynamic PD voltage, i.e. the dynamic generated light power, where the soliton step sits in Zone 2. 
Supplementary Information shows that Zone 2 covers the full range of detected voltages for all possible soliton steps in 50 times measurements.
Zone 1 indicates that the laser frequency is effectively blue-detuned, and Zone 3 indicates far red-detuned, neither of which allows soliton existence. 

To distinguish the soliton number, we use the backward tuning scheme by increasing the laser frequency \cite{Guo:16}. 
The soliton power trace probed by the PD is converted to a voltage trace that shows clear stair feature in Fig.~\ref{Fig:3}(b). 
Each step in the stair corresponds to a soliton state with different $N$. 
We measure the mean voltage $\overline{V}$ values of each step and classify them into different soliton number $N$. 
Figure \ref{Fig:3}(c) shows that, after repeating 20 times, a linear relation between the voltage and soliton number is found, i.e. $\overline{V} =a\times N+b$, with $a=0.0609$ and $b=0.0265$.

Previously, program-controlled single-soliton generation has been realized on integrated microresonators with repetition rates above 49 GHz \cite{WangX:21, Zhou:22} that are difficult to detect with normal fast photodetectors. 
Here, with calibrated zones and $(a,b)$ parameters, we demonstrate a program for deterministic generation of target $N-$soliton state in our Si$_3$N$_4$ microresonator of 19.97 GHz FSR, in addition to the single-soliton state. 
The program logic is shown in Fig.~\ref{Fig:3}(d) which consists two blocks. 
The first block (blue) is to land on a soliton step, and the second block (red) is to produce the target $N-$soliton state. 
Initially, the laser frequency scans from $\nu_s$ to $\nu_e$ ($\nu_s>\nu_e$) close to the resonance, such that the final PD voltage lands in Zone 3, as shown by the lightest blue trace in Fig.~\ref{Fig:3}(a). 
The scanning window is then translated by $\Delta \nu$, i.e. the laser re-scanned from $\nu_s+\Delta \nu$ to $\nu_e+\Delta \nu$. 
After that, if the PD voltage lands in Zone 1, meaning that the laser frequency is blue-detuned and the soliton state is not yet reached, then the system is re-initiated. 
If the PD voltage still lands in Zone 3, the laser frequency scanning window is further translated by $\Delta \nu$.
The process continues until the PD voltage lands in Zone 2, signalling that a soliton state is reached, as the darkest blue trace showing in Fig.~\ref{Fig:3}(a). 

When a soliton state is reached, the program's second block starts. 
Using $(a,b)$ parameters and the PD voltage, the current soliton number is estimated by $(V-b)/a$. 
For a target soliton number $N$, if $(V-b)/a<N-0.5$, the target $N-$soliton generation fails and the system re-initiated.  
If $(V-b)/a>N+0.5$, the laser frequency is increased with step $\Delta \nu ^ \prime$ until that $(V-b)/a\in [N-0.5, N+0.5]$ where $N-$soliton state is successfully produced.
Parameters $(\Delta \nu , \Delta \nu ^ \prime)$ are self-adaptive modified to accelerate the sequence (see Supplementary Information). 

We repeatedly apply the program for automatic $N-$soliton generation for 100 times. 
The soliton number $N$ can be retrieved by inverse Fourier transform on the soliton's optical spectrum on OSA, as shown in Fig~\ref{Fig:4}(a). 
We verify the soliton number of every addressed $N=5,4,3,2,1-$soliton state. 
The statistic charts of time consumption for the $N=3,2,1-$soliton state addressing are shown in Fig.~\ref{Fig:4}(b). 
The success rates are 93\%, 99\% and 100\% for $N=3,2,1-$soliton production, respectively. 
Table.~\ref{tab:summary} summarizes the data of up to 5-soliton production. 
The existence of $N-$soliton state is monitored by sampling the PD voltage $V$ at 10 Hz rate.

\begin{table*}[t!]
\centering
\caption{Statistics of success rate and time consumption to produce $N-$soliton states.}
\label{tab:summary}
\begin{tabular}{|c|c|c|c|c|c|}
\hline
Soliton number & Success rate & Probability & Probability & Mean time & Shortest time  \\
$N$ &  & < 15 s & < 20 s & (s) & (s)  \\
\hline
1 & 100\% & 59\% & 96\% & 14.6 & 10.6 \\
\hline
2 & 99\% & 65\% & 88\% & 14.7 & 10.3 \\
\hline
3 & 93\% & 63\% & 96\% & 14.3 & 10.1 \\
\hline
4 & 96\% & 45\% & 72\% & 17.5 & 9.9 \\
\hline
5 & 95\% & 47\% & 80\% & 16.7 & 2.8 \\
\hline
\end{tabular}
\end{table*}

In conclusion, we have demonstrated a novel experimental approach to avoid photo-thermal effect commonly encountered in the generation of dissipative Kerr solitons in optical microresonators. 
The approach is based on phase-modulation of the CW pump, where the generated blue sideband synergizes with the carrier pump and thermally stabilizes the microresonator. 
The sideband provides immediate heating when the intra-cavity power transits to a soliton state and the microresonator experiences a sudden cooling. 
Using a 19.97-GHz-FSR Si$_3$N$_4$ microresonator, we successfully demonstrate soliton step extension by 8 times with the sideband heating. 
Furthermore, we have developed a program to automatically, deterministically and quickly address to target $N-$soliton state, including the single-soliton state. 
We show near 100\% success rate and as short as 10 s time consumption for $N-$soliton state addressing. 

While demonstrated on Si$_3$N$_4$, our method is also useful for soliton generation in other platforms suffering from stronger photo-thermal effect. 
The program-controlled automatic addressing of $N-$soliton state can be critical for several applications in microwave photonics and frequency comb metrology. 
Together, our method facilitates a wider applications of soliton microcomb systems out of laboratory environments.

\medskip
\begin{footnotesize}

\noindent \textbf{Funding Information}: 
J. Liu acknowledges support from the National Natural Science Foundation of China (Grant No.12261131503), 
Shenzhen-Hong Kong Cooperation Zone for Technology and Innovation (HZQB-KCZYB2020050), 
and from the Guangdong Provincial Key Laboratory (2019B121203002). 
Y.-H. L. acknowledges support from the China Postdoctoral Science Foundation (Grant No. 2022M721482). 
Hairun Guo acknowledges support from the National Natural Science Foundation of China (Grant No.11974234).

\noindent \textbf{Acknowledgments}: 
Silicon nitride chips were fabricated by Qaleido Photonics. 
We thank Yuan Chen, Sanli Huang and Zhongkai Wang for assistance in the experiment, Suwan Sun for assistance in simulation, Jijun He for the fruitful discussion. 

\noindent \textbf{Author contributions}: 
W. S. and J. Liu conceived the experiment.
H. Z., W. S., H. W., B. S. and J. Long performed the experiment, assisted by S. M.. 
X. D., R. C. and Y.-H. L. performed the simulation, supervised by H. G..
C. S. assisted in fabricating Si$_3$N$_4$ chips.
W. S., Y.-H. L., B. S. and J. Liu prepared the manuscript.
J. Liu supervised the project and managed collaboration.

\noindent \textbf{Disclosures}: 
W. S., J. Long and J. Liu filed a patent application related to this work. 
Others declare no conflicts of interest. 

\noindent \textbf{Data Availability Statement}: 
The code and data used to produce the plots within this work will be released on the repository \texttt{Zenodo} upon publication of this preprint.

\end{footnotesize}

%

\end{document}


\title{Programmable access to microresonator solitons with modulational sideband heating}

\author{Huamin Zheng}
\affiliation{College of Electronics and Information Engineering, Shenzhen University, Shenzhen 518000, China}
\affiliation{International Quantum Academy, Shenzhen 518048, China}

\author{Wei Sun}
\email[]{sunwei@iqasz.cn}
\affiliation{International Quantum Academy, Shenzhen 518048, China}

\author{Xingxing Ding}
\affiliation{Key Laboratory of Specialty Fiber Optics and Optical Access Networks, Shanghai University, Shanghai 200444, China}

\author{Haoran Wen}
\affiliation{International Quantum Academy, Shenzhen 518048, China}
\affiliation{School of Science and Engineering, CUHK(SZ), Shenzhen 518100, China}

\author{Ruiyang Chen}
\affiliation{International Quantum Academy, Shenzhen 518048, China}
\affiliation{Shenzhen Institute for Quantum Science and Engineering, Southern University of Science and Technology, Shenzhen 518055, China}

\author{Baoqi Shi}
\affiliation{International Quantum Academy, Shenzhen 518048, China}
\affiliation{Department of Optics and Optical Engineering, University of Science and Technology of China, Hefei 230026, China}

\author{Yi-Han Luo}
\affiliation{International Quantum Academy, Shenzhen 518048, China}
\affiliation{Shenzhen Institute for Quantum Science and Engineering, Southern University of Science and Technology, Shenzhen 518055, China}

\author{Jinbao Long}
\affiliation{International Quantum Academy, Shenzhen 518048, China}

\author{Chen Shen}
\affiliation{International Quantum Academy, Shenzhen 518048, China}

\author{Shan Meng}
\affiliation{College of Electronics and Information Engineering, Shenzhen University, Shenzhen 518000, China}

\author{Hairun Guo}
\affiliation{Key Laboratory of Specialty Fiber Optics and Optical Access Networks, Shanghai University, Shanghai 200444, China}

\author{Junqiu Liu}
\email[]{liujq@iqasz.cn}
\affiliation{International Quantum Academy, Shenzhen 518048, China}
\affiliation{Hefei National Laboratory, University of Science and Technology of China, Hefei 230088, China}
\maketitle

\section{Simulation on soliton step extension}

As mentioned in the main text, a modified Lugiato-Lefever Equation (LLE) model to simulate soliton formation is considered as below,
\begin{equation}
    \begin{aligned}
        \frac\partial{\partial t}A(t,\varphi)&=-\frac \kappa2A(t,\varphi)-i[\delta_{\omega}+g_{\omega}|A|^2+\mathcal{D}_{\omega}+T]A(t,\varphi) +\sqrt{\kappa_{\text{ex}}}A_{\text{in}}\\
        \frac{\partial T}{\partial t} &=\frac{KT}{\tau_{\text{r}}}|A(t,\varphi)|^2-\frac T{\tau_{\text{r}}}
    \end{aligned}
\end{equation}
with
\begin{equation}
    \begin{aligned}
        \kappa & =\kappa_0 + \kappa_{\text{ex}}  \\
        g_{\omega}& =2\pi\frac{\omega_{0}cn_{2}}{n_{0}^{2}V_{\mathrm{eff}}}\cdot (1-\frac{D_{1}}{\omega_{0}}i\partial_{\varphi}) \\
        \mathcal{D}_{\omega}& =\sum_{m=2}m!^{-1}D_{m}(\mu_{0})(-i\partial_{\varphi})^{m}  \\
        A_{\text{in}}&=\sqrt{P_{\text{in}}}(1+\delta_{\text{m}}\text{exp}(+i2\pi f_{\text{m}}t))
    \end{aligned}
\end{equation}

where $D_m(\mu_0)$ represents the $m^\text{th}$-order dispersion term at the $\mu_0$ resonance mode,
$\omega_0$ is the cavity resonance frequency at the $\mu_{0}$ resonance mode,
$n_2$ is the nonlinear refractive index,
$n_0$ is the refractive index,
$V_\text{eff}$ is the effective mode volume.

We experimentally characterize the Si$_3$N$_4$ microresonator using a vector spectrum analyzer \cite{Luo:23} with measured parameters as 
$\kappa/2\pi=27.8$ MHz ($\kappa_\text{ex}/2\pi=7.4$ MHz and $\kappa_0/2\pi=20.4$ MHz), 
$D_1/2\pi=19.97$ GHz, 
$D_2/2\pi=44.79$ kHz, 
$D_3/2\pi=-7.38$ Hz. 
$n_0=1.846$ and
$V_\text{eff}=9.05\times 10^{-15}$ m$^3$ are obtained from COMSOL simulation.
$n_2=2.4\times 10^{-19}$ m$^2$/W from the literature\cite{Ikeda:08}.
The pump wavelength is $\lambda_\text{p}=1555.7$ nm ($\omega_0/2\pi=c/\lambda_\text{p}$) and the input power is $P_\text{in}=200$ mW on the chip.

In a Si$_3$N$_4$ microresonator with neither the photo-thermal effect nor the sideband, i.e. $T=0$ and $\delta_\text{m}=0$, the intra-cavity power features a small triangle shape and a soliton step as long as $20(\kappa/2\pi)$, as shown in Supplementary Fig.~\ref{FigS:simulation} left. 
When the photo-thermal effect is considered, the triangle waveform extends and the soliton step shrinks as shown in Supplementary Fig.~\ref{FigS:simulation} middle, with thermally induced resonance shift $KT=2.2\times 10^{11}\cdot \hbar \omega_{0} \kappa$.
The intra-cavity temperature drops immediately when the soliton state is reached on the effectively red-detuned side.
Further, when the blue sideband is introduced in the simulation, the intra-cavity temperature change can be compensated by the sideband, and the soliton step is extended dramatically, as shown in Supplementary Fig.~\ref{FigS:simulation} right, with $\delta_\text{m}=0.1$ and $f_\text{m}=695$ MHz.
The simulation in the main text Fig.~1 (b) shares the same parameters here except for $\lambda_\text{p}=1556.9$ nm, $P_\text{in}=400$ mW, $KT=5.0\times 10^{11}\cdot \hbar \omega_0 \kappa$, $\delta_\text{m}=0.26$ and $f_\text{m}=1.42$ GHz. 

From simulation with different parameters, it shows that an additional sideband can compensate temperature change and effectively extend a short soliton step.

\begin{figure*}[h!]
\renewcommand{\figurename}{Supplementary Figure}
\centering
\includegraphics{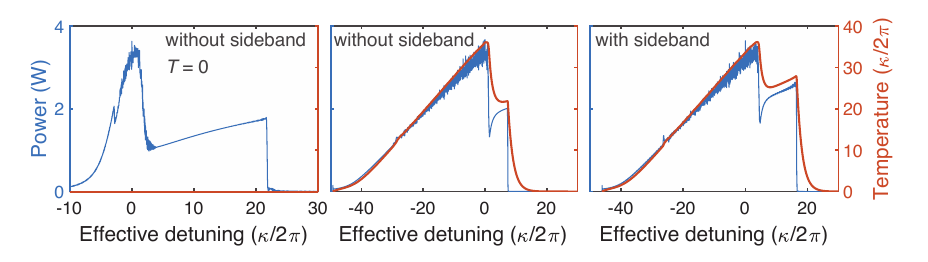}
\caption{
\textbf{Numerical simulation.}
Evolution of intra-cavity power (blue lines) and temperature (red lines).
Left: without the photo-thermal effect and sideband heating. 
Middle: with the photo-thermal effect and without sideband heating. 
Right: with the photo-thermal effect and sideband heating.
}
\label{FigS:simulation}
\end{figure*}

\section{Characterization of the Si$_3$N$_4$ microresonator}
For soliton generation, anomalous group velocity dispersion (GVD) is required.  
The microresonator's integrated dispersion is defined as
\begin{equation}
    \begin{aligned}
        D_{\mathrm{int}}(\mu)& =\omega_{\mu}-\omega_{0}-D_{1}\mu   \\
        &=D_{2}\mu^{2}/2+D_{3}\mu^{3}/6+D_{4}\mu^{4}/24+\dots
    \end{aligned}
    \label{Eq:dispersion}
\end{equation}
where $\omega_\mu/2\pi$ is the $\mu^\text{th}$ resonance frequency relative the the reference resonance frequency $\nu_0=\omega_0/2\pi$, 
$D_1/2\pi$ is the FSR of the microresonator, 
$D_2/2\pi$ describes GVD, $D_3$, $D_4$ and so on represent higher-order dispersion. 

We use a vector spectrum analyzer \cite{Luo:23} to characterize the Si$_3$N$_4$ microresonator. 
The data measured from our Si$_3$N$_4$ microresonator is shown in Supplementary Fig.~\ref{FigS:chip}(a), and parameters are extracted from fitting the data with Eq.~\ref{Eq:dispersion}.
$D_2=44.79$ kHz is positive meaning an anomalous GVD. 
Dispersion terms higher than $D_3$ are not considered in the above simulation.

Supplementary Fig.~\ref{FigS:chip}(b) shows a typical $\text{TE}_{00}$ resonance profile with Lorentzian fit around $\lambda_\text{p}=1555$ nm. 
The resonance is under-coupled ($\kappa_\text{ex}<\kappa_0$) with $\kappa_0/2\pi=20.9$ MHz. 
A histogram of $\kappa_0/2\pi$ from 990 fitted resonances shows the most probable value of $21$ MHz, corresponding to a statistical intrinsic quality factor of $Q_0=9.2\times 10^6$.

\begin{figure*}[h!]
\renewcommand{\figurename}{Supplementary Figure}
\centering
\includegraphics{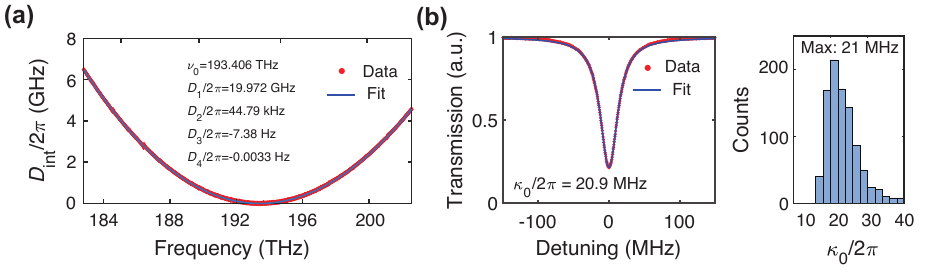}
\caption{
\textbf{Characterization of the Si$_3$N$_4$ microresonator.}
\textbf{(a)}. 
Experimentally measured dispersion profile $D_\text{int}$ of the Si$_3$N$_4$ microresonator. 
The red dots are measured data and the blue line is the fitting curve, with each extracted dispersion parameters marked.
\textbf{(b)}. 
Data of microresonator loss. 
Left: one representative resonance (red dots) and its Lorentzian fit (blue curve). 
Right: statistical analysis of 990 retrieved $\kappa_0/2\pi$ from this Si$_3$N$_4$ microresonator. 
}
\label{FigS:chip}
\end{figure*}

\section{Determination of Zone 2}

To generate a soliton state, the pump laser frequency is scanned from the effectively blue-detuned side to the red-detuned side with soliton step features, as shown in Supplementary Fig.~\ref{FigS:step}.
Experimentally, the generated soliton number in each scan is random, leading to discrete soliton step altitude and length.
We repeat the scans for 50 times with sideband heating and assemble all the soliton step altitudes to one Zone, i.e. the Zone 2 in the main text.
Power falling in Zone 2 serves as the criterion for successful soliton state excitation.

\begin{figure*}[h!]
\renewcommand{\figurename}{Supplementary Figure}
\centering
\includegraphics{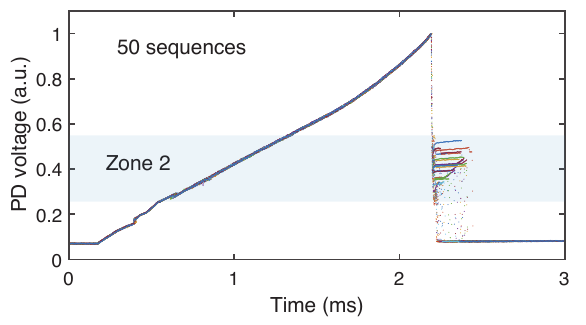}
\caption{
\textbf{Determination of Zone 2.}
Traces with different color represents different scan trials. 
The shallow blue area is labeled as Zone 2 where all the soliton steps fall in the 50 scans. 
}
\label{FigS:step}
\end{figure*}
\pagebreak


\pagebreak

\section{Sequence acceleration}

To accelerate the sequence for $N-$soliton state addressing, frequency translation $\Delta\nu$ and $\Delta\nu'$ are applied according to the ``distance'' between the current state to the target state. 
To reach a soliton state, the time distance $\Delta t$ is extracted between the time at peak PD voltage and the time at laser frequency $\nu_{e}$, as shown in Supplementary Fig.~\ref{FigS:acceleration}(a). 
If $\Delta t>\Delta t_\text{c}$ ($\Delta t_\text{c}$ is set as a critical time distance), the frequency window $(\nu_{s},\nu_{e})$ translates with $\Delta\nu=\Delta\nu_1= k \Delta t$. 
In the next frequency scanning sequence, the profile of PD voltage moves to right with soliton step approaching to the end frequency $\nu_{e}$ as described in the main text, and the time distance $\Delta t$ gets short.
If $\Delta t<\Delta t_\text{c}$, the frequency window $(\nu_{s},\nu_{e})$ translates with $\Delta\nu=\Delta\nu_2= C_0$ where $C_0$ is a constant value.
The algorithm flowchart is shown in Supplementary Fig.~\ref{FigS:acceleration}(c).
It can be seen that the frequency translation $\Delta\nu$ changes in a self-adaptive way.
The frequency moves fast when the time distance is large and slows down its speed when the time distance becomes small.
When the time distance comes to a critical point, i.e. $\Delta t=\Delta t_\text{c}$, the frequency moves in a constant small speed which guarantees not missing soliton steps.
By adjusting values of $\Delta t_\text{c}$ and $k$, a soliton state can be generated in a few seconds, as shown in Supplementary Fig.~\ref{FigS:acceleration}(b). 

To address to a target $N-$soliton state, acceleration is also applied, as shown in Supplementary Fig.~\ref{FigS:acceleration}(d). 
The ``distance'' between $(V-b)/a$ and $N$ is accounted for setting the frequency translation. 
We set $\Delta\nu' =\Delta\nu_1'$ when $(V-b)/a > N+1.5$, and  $\Delta\nu'=\Delta\nu_2'$ elsewise ($\Delta\nu_1' > \Delta\nu_2'$). 
The two-stage acceleration enables $N-$soliton state addressing in as short as 10 s.

\begin{figure*}[h!]
\renewcommand{\figurename}{Supplementary Figure}
\centering
\includegraphics{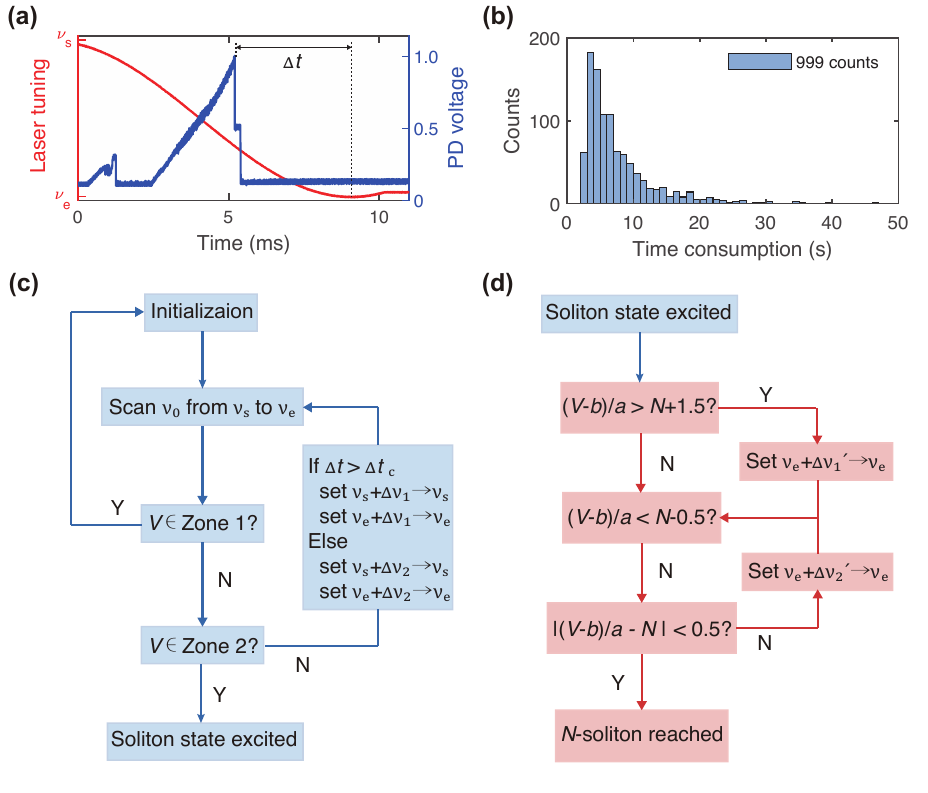}
\caption{
\textbf{Sequence acceleration.}
\textbf{(a)}. 
Frequency scanning (red curve) profile, and PD voltage (blue) corresponding to the generated light in microresonator. 
Dashed lines mark the $\Delta t$ between the time at peak PD voltage and the time at frequency $\nu_{e}$.
\textbf{(b)}. 
Statistics on time consumption to generate a soliton state.
\textbf{(c)}. 
Time sequence to accelerate soliton state generation.
\textbf{(d)}. 
Time sequence to accelerate the target $N-$soliton state addressing.
}
\label{FigS:acceleration}
\end{figure*}
\pagebreak


\section*{Supplementary References}
\bigskip
%